\documentclass[11pt,epsfig]{article} 
\usepackage{epsfig}
\usepackage[small,hang]{caption}
\usepackage{cite}  
\usepackage{color}
\def\epsbo{\mbox{\boldmath $\epsilon$}}

\newcommand{\beq}{\begin{equation}}
\newcommand{\eeq}{\end{equation}}
\newcommand{\bea}{\begin{eqnarray}}
\newcommand{\eea}{\end{eqnarray}}       
\newcommand{\beqn}{\begin{equation}} 
\newcommand{\eeqn}{\end{equation}}
\newcommand{\mbf}[1]{\mbox{\boldmath $#1$}}
\newcommand{\ba}{k_1^*}
\newcommand{\bb}{k_2^*}
\newcommand{\bh}{\bar{h}}
\newcommand{\bk}{\mbf{k}}

\newcommand{\al}{\alpha}

\newcommand{\asb}{\bar{\alpha}_s}

\begin{document}
\begin{titlepage}
\noindent
DESY 03-053 \hfill
ISSN 0418--9833\\

\vspace*{2.cm}
\begin{center}
  \begin{LARGE}
    {\bf QCD Corrections to Electroweak Vector Boson \\ Scattering at Small
    Scattering Angles}\\
  \end{LARGE}
  \renewcommand{\thefootnote}{\fnsymbol{footnote}}
  \renewcommand{\thefootnote}{\arabic{footnote}}
  \setcounter{footnote}{0}
  \vspace{2.5cm}
  \begin{Large}
    {Kriszti\'an Peters$^{(a)}$\footnote{\noindent
        email: krisztian.peters@desy.de}
%  $\quad$ \\
\renewcommand{\thefootnote}{\fnsymbol{footnote}}
\renewcommand{\thefootnote}{\arabic{footnote}}\hspace{-0.5cm}
 and Gian Paolo Vacca$^{(b)}$}\footnote{email:
        vacca@bo.infn.it  $\quad$       }     \\
  \end{Large}
  \vspace{0.3cm}
  \textit{$^{(a)}$II.\ Institut f\"ur Theoretische Physik,
    Universit\"at Hamburg,\\ Luruper Chaussee 149,
    D-22761 Hamburg}  \\
 \vspace{0.2cm}
  \textit{$^{(b)}$ Dipartimento di Fisica - Universit\`a di Bologna and
INFN - Sezione di Bologna,\\
via Irnerio 46, 40126 Bologna, Italy}
\end{center}
\vspace*{1.5cm}
\begin{abstract}
We investigate the role of a certain class of QCD corrections to
electroweak vector boson scattering at small scattering angles and
large energies.
These are present since, from the perturbative analysis,
the vector bosons may dissociate into quark-antiquark
pairs giving rise to colour dipoles interacting through gluon exchanges.
After the computation of the vector boson impact factors, 
we present expressions for the lowest order QCD scattering amplitude
and for the leading logarithmic BFKL amplitude.
Particularly we discuss numerical results for the process
$\gamma\gamma\to ZZ$. The QCD corrections to the cross
section resulting from the interference with the electroweak
ones are estimated and compared with the leading pure electroweak part.
Corrections resulting from the leading log BFKL amplitude 
are of the order of few percent already at the $0.5 - 1$ TeV energy range.
\end{abstract}
\end{titlepage}

\section{Introduction}
The search for physics beyond the Standard Model (SM) and the
determination of the crucial and still not understood mechanism
of the electroweak symmetry breaking in the SM itself are major subjects
of investigation in particle physics. 
There is the hope to increase the understanding with the help of the
next generation of supercolliders, and major efforts in this direction have
currently been made. First insights are expected from the
future large hadron collider (LHC) runs.
On the other hand high precision measurements at the Next Linear
Collider (NLC) are considered to be of great importance,
with features complementary to LHC.
Next Linear Collider is a generic name for a $e^+e^-$ machine
operating in the energy regime up to 1 TeV,  
providing a very clean environment, and making available observables for
many processes at an accuracy better than the percent level.
In addition to the $e^+e^-$ collider mode, one may
have the capability of running the NLC in a $\gamma\gamma$ collision
mode via Compton backscattering of laser photons off the linear
collider electrons. 

One of the important processes one may consider in a linear collider
is the scattering of vector bosons.
Indeed, at the NLC the electron and positron bunches can radiate
vector bosons which interact thereafter. Diverse scenarios are
considered also in the Compton collider mode.
Each incoming photon may turn into
a virtual WW pair, followed by the scattering of one W of each pair
to end with WW or ZZ states~\cite{BBB}.
There are also the processes where initial photons scatter directly
into arbitrary vector bosons. 

Vector boson scattering is related to the open question of mass generation.
Since the longitudinal modes of the massive vector bosons reflect the nature 
of the Goldstone modes of the unbroken theory, these may give important
informations about the nature of electroweak symmetry breaking.
Without the discovery of a scalar Higgs particle 
the vector bosons have to interact strongly at TeV scales, in order to preserve
unitarity. In this scenario the vector boson scattering will be one
of the key processes to be studied in the nearest future~\cite{LQT}.

Another important question one may pose while searching for new physics is the
existence of anomalous triple and quartic vector boson couplings which
may affect vector boson scattering.
Contact interactions are regarded as a possible extension of the SM,
which can be well investigated with vector boson scattering~\cite{Anom}.
The natural order of magnitude of the anomalous couplings~\cite{B} is
small and, in order to eventually extract them, one needs to both
measure and compute theoretically the SM cross sections with a
precision better than $1\%$.

In this context the $\gamma\gamma\to ZZ$ process has a special role.
Since there is no perturbative contribution at the tree level, such a
process is sensitive to the particles and the new physics phenomena
which contribute through radiative corrections \cite{GLPR}.
This leads to an independent and complementary method to
the direct production of new particles. 
In the same process the detection of CP violating phases \cite{CP} or 
additional effects due to the exchange of Kaluza-Klein gravitons in
large extra-dimension scenarios \cite{ExtraD} have been discussed.

In summary, vector boson scattering processes may be seen as a central
tool to probe physics beyond the SM and to reveal the nature
of symmetry breaking. In order to exploit these features, a high
precision is needed, both on experimental and on theoretical side.
Calculations of these processes at the lowest
available order may not be accurate enough.
One loop corrections have been calculated for all those vector bosonic
processes~\cite{BosonKorr} where a tree level contribution is present.
For the $\gamma\gamma\to ZZ$ process, which starts at one loop in 
$\alpha_{EW}$, no higher order corrections are available yet.
The question naturally arise, if this accuracy is high enough to
disentangle new physics from the SM, providing a window to
non-accessible energy regimes.

The aim of this paper is to address the question if the QCD corrections
could play a role in vector boson scattering processes.
Naively this seems not to be the case, since they arise
only at higher loop corrections.
On the other hand, even if the lowest order non-suppressed 
QCD correction to the
amplitude for this process is smaller by $O(\al_s^2)$ w.r.t. the
electroweak one, in the cross section the interference among the two
can reach the percent level.
Moreover at high energies there are kinematical regimes wherein the
emission of soft gluons generates large logs in the energy which make
the QCD amplitude rising with energy.

The electroweak cross section of $VV\to VV$ (with $V=W,Z,\gamma$) 
gains its biggest contribution from small scattering angles, and
with rising scattering energy the cross sections are approaching a
constant in computations up to the one loop level.
This contribution is coming from diagrams where only spin-1
particles are exchanged in the $t$-channel.
At leading order and $O(\alpha^2_{EW})$ these exchanged particles
are only electroweak vector bosons.

The same configuration arises for QCD corrections with gluons in the
$t$-channel. The first gluonic contributions, which arise at the two
loop level, are suppressed since these gluons are always accompanied
by fermions in the $t$-channel.
The non-suppressed QCD contributions occur at higher orders.
These emerge when the vector bosons fluctuate into a
quark-antiquark pair and these dipoles interact through gluons.
At the lowest order, when two gluons are exchanged in the $t$-channel,
one obtains a contribution constant in energy while higher order
corrections provide large logarithms in the energy.
These contributions cannot be neglected at large energies.
The extreme attitude is to consider a resummation described by the
BFKL equation ~\cite{BFKL} which gives an upper bound estimate of
these effects.

We study in detail these QCD corrections for the
reaction $\gamma\gamma\to ZZ$, whose cross section 
up to now was only studied at leading order
(one loop in the electroweak sector) by G. Jikia ~\cite{Jikia}.
His results show that the cross section
is approaching a constant for high energies and its main contribution
comes from the helicity conserving amplitudes at small scattering
angles.
This is exactly the kinematical region where the above discussed QCD
contributions have their biggest influence.

We begin with the analytic calculations for an arbitrary bosonic
process. In the next section the two gluon
exchange amplitude is computed, including the calculation of the
non-forward impact factor. Since we consider the high energy limit
the amplitude factorise into two independent impact factors, built from
the bosonic wave functions which describe the probability that a
vector Boson fluctuates into a quark anti-quark pair.
In the third section the cross section of the resummed gluonic
contribution is presented, i.e. the BFKL amplitude.
In order to give some numerical estimates, we concentrate on 
the definite process $\gamma\gamma\to ZZ$.
In the fourth section we calculate the EW and QCD parts of the
amplitude in the Regge limit. Next we evaluate the corrections numerically.
Because of the strong scale dependence we
can give only estimates of the corrections, which we find to be of
the order of a few percent. Finally we present our conclusions.
%We consider our result as a warning against interpreting
%percent level deviations from the EW calculations at small angles 
%as signals of new physics.
Our result suggest that, in particular in the small angle region,
possible deviations from EW calculations cannot be interpreted as
signals of new physics without taking into account the QCD~corrections.
%%%%%%%%%%%%%%%%%%%%%%%%%%%%%%%%%%%%%%%%%%%%%%%%%%%%%%%%%%%%%%%%%%%%%%%%%%
\begin{figure}[t]
  \begin{center} 
     \epsfig{file=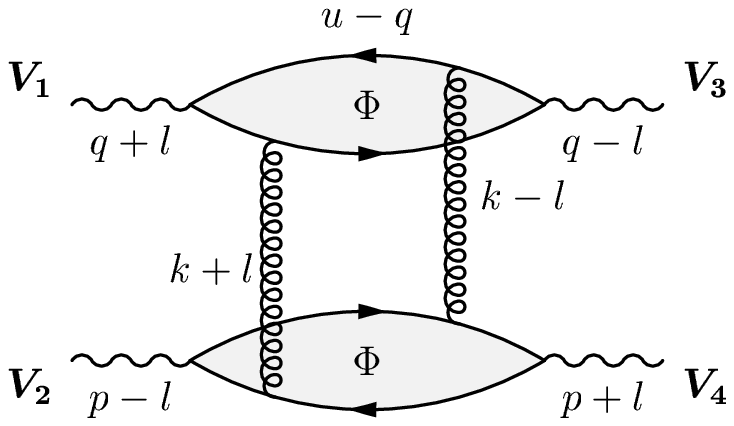,bbllx=198pt,bblly=582pt,bburx=408pt,bbury=728pt,clip=}
     \caption{The two gluon exchange of the $VV\to VV$
        elastic cross section.}
     \label{2gl}
\end{center}
 \end{figure}
%%%%%%%%%%%%%%%%%%%%%%%%%%%%%%%%%%%%%%%%%%%%%%%%%%%%%%%%%%%%%%%%%%%%%%%%%%%%%

\section{Impact factors and lowest order QCD contribution}
At high energies the lowest order leading QCD contribution to the vector boson
scattering is pictured in Fig.\ref{2gl} where the lowest
order diagrams with two-gluon exchange are illustrated.
The vector boson fluctuation into a quark anti-quark pair is described
by the product of the two boson wave functions.
This quantity is called the (non-forward) vector boson impact factor.
Due to the high-energy factorization one can
compute first the impact factors and then integrate these with the gluon
propagators. In the high-energy limit, corresponding to $s$ going to infinity
and $t$ fixed (Regge limit), we calculate first the $s$-channel
discontinuity and restore after the amplitude through dispersion
relations. Thus the intermediate quark lines inside the impact factors
are taken on-shell. This procedure and the high-energy factorization
simplifies the calculation significantly.
The $s$-channel discontinuity of the scattering amplitude takes the form:
\begin{eqnarray}
\label{2glampl}
Disc_s A=is \int
\frac{d^2{\bf k }}{(2\pi)^3}\;\frac{\Phi(q,k,l)}{{\bf (k+l)}^2 }
\frac{\Phi(p,-k,-l)}{{\bf (k-l)}^2 } \, ,
\end{eqnarray}
where $\Phi$ is the impact factor and the boldface
marked letters are the transverse part of the euclidean momenta. 
As usual, the integration of the two longitudinal components of the loop
momenta are absorbed into the definition of the impact factors.
The remaining integrals may be computed numerically.

The momenta of the incoming bosons are $q+l$ and $p-l$, while the
outgoing bosons have momenta $q-l$ and $p+l$.
We use Sudakov variables with the light cone vectors $p^\prime =p + y
q^\prime$, $y\simeq -p^2/2p^\prime\cdot q^\prime$ and
$q^\prime=q+xp^\prime$, $x\simeq -q^2/2p^\prime\cdot q^\prime$, 
while $s=(p+q)^2\simeq 2 p^\prime\cdot q^\prime$ denotes as usual the
squared center of mass energy of the $VV$ scattering process.
Then one parametrizes any 4-momentum $u$ as
\beq
  \label{eq:sud}
  u= \alpha_u q^\prime + \beta_u p^\prime + u_\perp \, .
\eeq
The impact factor for incoming and outgoing photons has been
calculated recently \cite{peters} and since we adopt the momentum
assignment which was given there, we refer the reader to this paper
for further details of the calculation.
\newpage
The incoming and outgoing bosons are characterized by the invariants
$Q_+^2=-(q+l)^2$,~$Q_-^2=-(q-l)^2$ and by similar expressions
associated to the other two bosons involved in the
process. For the full process the Mandelstam variable $t=(2l)^2$ has
to be considered. In the case of on-shell bosons the virtuality
reduces to the boson mass, i.e. $Q^2=-M_V^2$.

We remind that in the limit of Regge kinematics one can neglect terms
which are suppressed by powers of $t/s$ and $Q_{\pm}^2/s$, leading to
important simplifications.
Let us consider the polarization vectors for the upper incoming and
outgoing bosons.
Their longitudinal polarization in terms of the Sudakov
variables are given by
\begin{equation}
  \label{eq:photl}
   \epsilon_L(q\pm l)\,=\,\frac1{|Q_\pm|}\left[(1\pm\alpha_l)\, q^\prime
  +\left( x\mp\beta_l-\frac{2 l^2_\perp}{s}\right) p^\prime
\pm l_\perp\right]\,  ,
\end{equation}
while the transverse polarisation vectors are 
\begin{equation}
  \label{eq:phott}
   \epsilon_T^{(h)} (q\pm l)\,=\,\epsilon^{(h)}_\perp
  \, \pm \,
  \frac{2
  l_\perp\cdot\epsilon^{(h)}_\perp}{s}\,\,(q^\prime-p^\prime\pm l_\perp)\,,
\end{equation}
where $h=\pm$ denotes two helicity states and 
\begin{eqnarray} \label{eq:hel}
 \epsilon^{(h)}_\perp\, =\, \frac1{\sqrt 2}\,(0,1,\pm\, i,0)\,  .
\end{eqnarray}
Again, for the bosons appearing in the lower impact factor the same
expression holds with an exchange of $q$ and $p$ and $l \to -l$.

A difference between the photon and the other bosons is given by the
chiral coupling to the quarks of the latters. Consequently extra terms
in the traces appear and the ones proportional  to odd powers of 
$\gamma^5$ do not contribute because of charge conjugation
invariance. Thus, regarding the impact factors, the $W$ and $Z$ bosons 
behave both in the same way as the
photon, i.e. we can use the result from \cite{peters}, and
only the magnitude of the coupling to the quarks depends on the nature
of the boson.

The impact factor where both of the bosons have longitudinal 
polarisations takes the form:
\begin{eqnarray} \label{LL}
&&\hspace{-2cm}\Phi_{LL}(q,k,l)=  \alpha_s\sqrt{N_c^2-1} \, \sum_f {\it
  C}_f\,
4\sqrt{4\pi} \; |Q_+||Q_-|\int_0^1 d\alpha\int \frac{d^2{\bf u}}{(2\pi)^2} \times
\\ \nonumber &&\hspace{4cm}\times\alpha ^2 (1-\alpha )^2 
\left( \frac1{D_1^+}-\frac1{D_2^+}\right)\left( \frac1{D_1^-}-
\frac1{D_2^-}\right)\, , 
\end{eqnarray}
while for transverse polarized bosons we have:
\begin{eqnarray} \label{TT}
&& \hspace{-1cm}
\Phi_{TT}^{(ij)}(q,k,l) \, =\, \alpha_s\, \sqrt{N_c^2-1} \sum_f {\it
  C}_f\; \sqrt{4\pi}\int_0^1 d\alpha \int \frac{d^2 {\bf u}}{(2\pi)^2}
\\ \nonumber
&& \times\,
\left\{ -4\alpha(1-\alpha)\,\,\, \epsbo_i \cdot
\left( \frac{{\bf N_1^+}}{D_1^+}-\frac{{\bf N_2^+}}{D_2^+}\right)
\left( \frac{{\bf N_1^-}}{D_1^-}-\frac{{\bf N_2^-}}{D_2^-}\right) \cdot
\epsbo_j^* \right. \\ \nonumber
&& \left. +\,
\epsbo_i \cdot  \epsbo_j^*
\left[
\left( \frac{{\bf N_1^+}}{D_1^+}-\frac{{\bf N_2^+}}{D_2^+}\right)\cdot
\left( \frac{{\bf N_1^-}}{D_1^-}-\frac{{\bf N_2^-}}{D_2^-}\right)
+ m_f^2\, \left( \frac1{D_1^+}-\frac1{D_2^+}\right)
\left( \frac1{D_1^-}-\frac1{D_2^-}\right) \right] \right\}\, .
\end{eqnarray}

Finally, if the incoming boson is transverse polarized while the outgoing one
is longitudinal, the impact factor reads:
\begin{eqnarray} \label{LT}
&& \hspace{-1cm}\Phi_{LT}^{(j)} (q,k,l)= \alpha_s\sqrt{N_c^2-1}
\, \sum_f {\it C}_f\, 2 \sqrt{4\pi} \; |Q_+| \int_0^1 d\alpha 
\int \frac{d^2 {\bf u}}{(2\pi)^2}  \times \\ \nonumber 
&&\hspace{3.5cm}
\times\, \alpha (1-\alpha )(1-2\alpha)  
\left( \frac1{D_1^+}-\frac1{D_2^+}\right)\left( \frac{{\bf N_1^-}}{D_1^-}-
\frac{{\bf N_2^-}}{D_2^-}\right)\cdot{\bf \epsilon_i^*}\, ,
\end{eqnarray}
and viceversa for the incoming boson longitudinal and the outgoing one
transverse, one has
\begin{eqnarray} \label{TL}
&& \hspace{-1cm}\Phi_{TL}^{(i)} (q,k,l)=\alpha_s\sqrt{N_c^2-1}
\,  \sum_f {\it C}_f\, 2 \sqrt{4\pi} \; |Q_-|\int_0^1 d\alpha
\int \frac{d^2 {\bf u}}{(2\pi)^2} \times \\ \nonumber
&&\hspace{3.5cm}\times\, \alpha  (1-\alpha )
(1-2\alpha)\; {\bf \epsilon_i}
\cdot\left( \frac{{\bf N_1^+}}{D_1^+}-\frac{{\bf N_2^+}}{D_2^+}\right)
\left( \frac1{D_1^-}-\frac1{D_2^-}\right)\, ,
\end{eqnarray}
where $\epsbo_j=1/\sqrt{2} (1,\pm i)$ are two-dimensional polarization
vectors corresponding to the two transverse polarizations $j=\pm$.
We have also used:
\begin{eqnarray}\nonumber
D_1^\pm&=&({\bf u}\pm (1-\alpha ) {\bf l})^2\,+\, \alpha (1-\alpha )
\,Q_\pm^2\,+\,m_f^2 \\ \nonumber
D_2^\pm&=&({\bf u-k}\,\mp\, \alpha{\bf l})^2\,\,+\,\alpha (1-\alpha )
\,Q_\pm^2\,+\,m_f^2
\\ \nonumber
{\bf N_1^\pm}&=& {\bf u}\pm (1-\alpha){\bf l} \\
{\bf N_2^\pm}&=&{\bf u-k}\mp\alpha{\bf l}\,.
\end{eqnarray}
Marked in boldface are the Euclidean
form of the transverse momenta, i.e. $u_\perp^2 = -{\bf u}^2<0$. 
The constant {\it $C_f$} is the product of the coupling of the incoming
and outgoing boson to the quarks of flavor $f$.
The impact factors have been written in a factorised form where the two
parts correspond to the boson wave functions, describing the
probability that a boson dissociates into a quark-antiquark pair.
In order to obtain the lowest order QCD amplitude, one has to
substitute these impact factors into  eq.(\ref{2glampl}) and integrate
over the gluon momenta.

The two helicity-flip impact factors are in general non-zero. They
vanish only for
forward scattering, where ${\bf l}=0$. This is
due to the different symmetry behaviour of the transverse and longitudinal 
boson wave functions. This property is
more trasparent in the coordinate space formulation, eq.(34) of \cite{peters}. 
Here the impact factor is written as a convolution of the
two boson wave functions and the dipol interaction. 
The dipole interaction is invariant under the
transformation of the dipole size vector $\bf{r\to -r}$, 
and so is the longitudinal
wave function, eq.(22) of \cite{peters}. In contrast to this 
the transverse wave
function, eq.(28-31) of \cite{peters}, is antisymmetric under this
transformation, since it is proportional to ${\bf r}$. 
The convolution of two transverse wave functions is
therefore symmetric under this transformation, whereas the convolution of
a transverse and a longitudinal one is antisymmetric, leading to the
vanishing result for ${\bf l}=0$. If ${\bf l}$ is non-zero, this
symmetry properties are broken, resulting in a non-vanishing
$\Phi_{TL}$ and $\Phi_{TL}$. 
%%%%%%%%%%%%%%%%%%%%%%%%%%%%%%%%%%%%%%%%%%%%%%%%%%%%%%%%%%%%%%%%%%%%%

\section{Leading log (BFKL) contribution}
At high energies and close to the forward region (Regge limit),
higher order contributions in $\alpha_s$ start to be important.
Diagrams of ladder topology, built with (non elementary) reggeized gluons,
can contribute up to the order $\alpha_s^n \ln^n (s/s_0)$.
On retaining the leading logarithmic (LL)
contributions and performing a resummation a rough estimate (overestimate) of
the total QCD corrections is obtained.
Nonetheless one expects to find the QCD contribution to the process
in the window constrained by the simple two gluon exchange on one side and
the resummed LL BFKL Pomeron exchange on the other.

\begin{figure}[t]
  \begin{center} \vspace{-0cm}
\epsfig{file=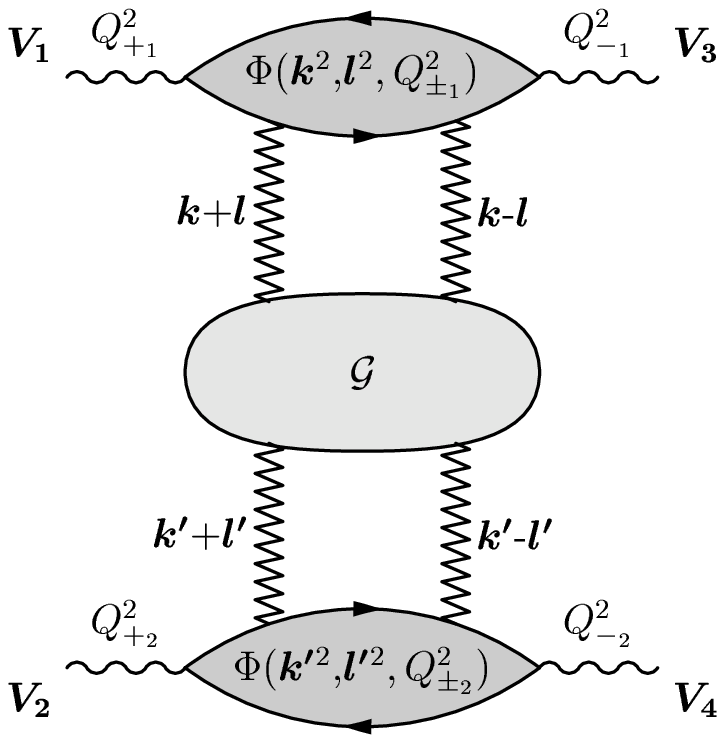,width=5cm,bbllx=192pt,bblly=491pt,bburx=420pt,
bbury=718pt,clip=}
     \caption{The BFKL Pomeron exchange in the $VV\to VV$
        elastic cross section.}
     \label{bfklampl}
\end{center}
 \end{figure}

This possibility is related to the property of the factorisation
of the QCD scattering amplitude, in the Regge limit, in terms of the impact
factors associated to the external particles and the BFKL Green's function,
as illustrated in Fig.~\ref{bfklampl}. 

 %%%%%%%%%%%%%%%%%%%%%%%%%%%%%%%%%%%%%%%%%%%%%%%%%%%%%%%%%%%%%%%%%%%%%%%%%%%%
%%%%%%%%%%%%%%%%%%%%%%%%%%%%%%%%%%%%%%%%%%%%%%%%%%%%%%%%%%%%%%%%%%%%%%%%%%%%%

For the physical process we want to study
the impact factors have been calculated in the preceeding section and
the discontinuity of the LL resummed amplitude is simply obtained by
inserting the non-forward BFKL Pomeron Green's function in the place
of the two gluon propagators. Indeed one has
\beq
Disc_s A=i s \int \frac{d^2 \mbf{k}}{(2\pi)^3}
\frac{d^2 \mbf{k}'}{(2\pi)^3}
\Phi(q, k, l)
G(y| \mbf{k}+\mbf{l}, \mbf{k}-\mbf{l}; \mbf{k}'+\mbf{l}, \mbf{k}'-\mbf{l})
\Phi(p, -k', -l),
\label{discA}
\eeq
where $y=\ln s/s_0$ is the rapidity variable and $s_0$ is a typical
scale in the process.
Clearly, for $\alpha_s \to 0$, when all rungs of the BFKL resummation
decouple, the Green's function reduces to describing the two gluon exchange,
which means that we use the following normalisation:
\beq
\lim_{\alpha_s \to 0}
G(y| \mbf{k}+\mbf{l}, \mbf{k}-\mbf{l}; \mbf{k}'+\mbf{l}, \mbf{k}'-\mbf{l})
=  \frac{(2\pi)^3}{(\mbf{k}+\mbf{l)}^2(\mbf{k}-\mbf{l})^2}
\delta^{(2)}(\mbf{k}-\mbf{k}')
\label{Gnorm}
\eeq
In the following we briefly give the ingredients needed to construct the
BFKL Green's function.
The eigenfunctions $E_{h,\bh}$ of the BFKL kernel, which appears in the
integral Bethe-Salpeter like equation resumming all the LL terms,
are well known in coordinate space, wherein their form is dictated by
conformal invariance. They are given by
\beq
E_{h,\bh}(\mbf{r}_{10},\mbf{r}_{20})=
\left(\frac{r_{12}}{r_{10}r_{20}}\right)^h
\left(\frac{r_{12}^*}{r_{10}^* r_{20}^*}\right)^{\bar{h}} \; ,
\label{pom_coord}
\eeq
with $\mbf{r}_{ij}=\mbf{r}_i-\mbf{r}_j$ etc.,
$h=(1+n)/2+i\nu$,
$\bar{h}=(1-n)/2+i\nu$ ($h^*=1-\bh$, $\bh^*=1-h$). Let us note that
standard complex notation for two-dimensional vectors is used on the
right-hand side. The conformal weights $h$, decomposed in $n$
(integer conformal spin) and $\nu$ (real), label the principal series
of unitary representations of the Moebius group.
Fourier transforming to momentum space (we use the Lipatov's convention which
assigns a $1/(2\pi)^2$ to any coordinate integration) one finds \cite{BBCV}:
\beq
\tilde{E}_{h\bh}(\mbf{k}_1,\mbf{k}_2)=
\int \frac{d^2\mbf{r}_1}{(2\pi)^2} \frac{d^2\mbf{r}_2}{(2\pi)^2}
E_{h,\bh}(\mbf{r}_1,\mbf{r}_2)
e^{i (\bk_1 \cdot \mbf{r}_1 + \bk_2 \cdot \mbf{r}_2 )}=  
C\Big(X(\mbf{k}_1,\mbf{k}_2)+
(-1)^nX(\mbf{k}_2,\mbf{k}_1)\Big).
\label{pom_mom}
\eeq
The coefficient $C$ is given by
\beq
C=\frac{(-i)^n}{(4\pi)^2}h\bh (1-h)(1-\bh)\Gamma(1-h)\Gamma(1-\bh).
\label{normpom}
\eeq
The functions $X$ in complex notation can be expressed in terms of
hypergeometric functions:
\beq
X(\mbf{k}_1,\mbf{k}_2)=\left(\frac{k_1}{2}\right)^{\bh-2}
\left(\frac{\bb}{2}\right)^{h-2}F\left(1-h,2-h;2;-\frac{\ba}{\bb}\right)
F\left(1-\bh,2-\bh;2;-\frac{k_2}{k_1}\right) \; .
\eeq
This analytic form does not contain any term of the type
$\delta^2( \mbf{k}_1)$ or $\delta^2( \mbf{k}_2)$ which are present in the 
coordinate representation eq.(\ref{pom_coord}).
The impact factor of a colorless external particle vanishes
for zero momentum of any gluon exchanged and therefore 
the delta-function type terms do not contribute.

The Pomeron intercept has the form $\alpha _{P}(0)=1+\chi(\nu,n)$, 
where 
\beqn
\chi(\nu,n)= \chi_h =
\bar{\alpha}_s\left( 2\psi(1)-\psi(\frac{1+|n|}{2}+i\nu)
                  -\psi(\frac{1+|n|}{2}-i\nu) \right)\, , \;
\bar{\alpha}_s=\frac{N_c\alpha_s}{\pi} \, ,
\eeqn
are the kernel eigenvalues corresponding to the eigenfunctions in equations
(\ref{pom_coord})-(\ref{pom_mom}).
We shall use the Pomeron Green's function in the momentum representation.
Starting from the form in coordinate representation~\cite{Lcft} one can perform a
Fourier transform~\cite{BBCV} and making use of the Casimir operator properties
of the M\"obius group, one finally obtains \cite{BRV}
for the non-amputated Green's function:
\beq
\tilde{G}_2(y|\bk_1,\bk_2;\bk_{1'}\bk_{2'})= (2\pi)^3 \int d \mu(h) \,
 e^{y \,\chi_h}
\, N_h \times (2\pi)^2 \,
\tilde{E}_{h,\bh}( \bk_1,\bk_2)
\tilde{E}_{h,\bh}^*( \bk_{1'}, \bk_{2'}) \;,
\label{G2mom}
\eeq
where we use the measure in the conformal weight space $\int d \mu(h) \, =
\sum_{n}\int d\nu$ with the following normalization factor $N_h$:
\beq
N_h= \frac{(\nu^2+n^2/4)}{[\nu^2+(n-1)^2/4][\nu^2+(n+1)^2/4]}.
\eeq
The $(2\pi)^3$ factor in front of the integral comes from the normalization
fixed by eq.(\ref{Gnorm}).

Let us finally rewrite the expression of the amplitude discontinuity
in eq.(\ref{discA}) by inserting in it the relation (\ref{G2mom}).
On defining the conformal weight representation for the impact factors
according to
\beq
\Phi_h(q,l)=\int \frac{d^2 \mbf{k}}{(2\pi)^3} \Phi(q, k, l)
\tilde{E}_{h,\bh}(\mbf{k}+\mbf{l}, \mbf{k}-\mbf{l}) \, ,
\label{conf_repr_phi}
\eeq
together with the primed $\Phi'_h(q,l)$ computed with the
conjugate eigenstate, 
one has
\beq \label{discAfinal}
Disc_s A=i s (2\pi)^5 \int d \mu(h) \,
 N_h \, e^{y \,\chi_h} \, \Phi_h(q,l) \Phi'_h(p,l)\, .
\eeq
We stress that due to high energy factorization the energy dependence
is encoded in the LL BFKL Green's function and the only other dependence
in the momenta of the incoming particles is through the invariant
external masses on which the impact factors depend.
In the last expression the dominant contribution for large $y$ can be easily
extracted with a saddle point analysis for $n=0$ around $\nu=0$ and
therefore reduces to
\beq \label{sad}
Disc_s A=i s (2\pi)^5 16 e^{y \, \chi_0} \frac{\sqrt{2\pi}}{(2 a y)^{3/2}}
\Phi_0(q,l) \Phi'_0(p,l) \, ,
\eeq
where $\chi_0=4 \ln(2) \asb$ is the leading BFKL intercept and
$a=14 \zeta(3) \asb$.

The forward limit presents a well known cusp behavior and it must
converge to the result obtained on using a forward BFKL Green's function
\beq
G_F(y|\bk,\bk')=(2\pi)^3 \int d \mu(h) \,  e^{y \,\chi_h}
\tilde{E}_{h,\bh}^{(F)}( \bk)
\tilde{E}_{h,\bh}^{(F)*}( \bk') \;,
\label{G2forw}
\eeq
which is written in the spectral represention with the forward BFKL amputated
eigenfunctions given by
$\tilde{E}_{h,\bh}^{(F)}( \bk)= 1/(\pi \sqrt{2}) k^{h-1}k^{* \, \bh -1}$.

The saddle point approximations can be carried on in a way similar to
the non forward case. We shall define
\beq
\Phi_h^{(F)}(q)=\int \frac{d^2 \mbf{k}}{(2\pi)^3}
\Phi(q, k, 0)\frac{1}{\mbf{k}^2}
\tilde{E}_{h,\bh}^{(F)}( \bk)\, ,
\label{forw_conf_repr_phi}
\eeq
as well as a $\Phi_h^{'(F)}(q)$ for the scalar product with the
conjugate Pomeron state and write the discontinuity in the forward
case as
\beq
Disc_s^{(F)} A=i s (2\pi)^3 e^{y \, \chi_0} \frac{\sqrt{2\pi}}{(2 a y)^{1/2}}
\Phi_0^{(F)}(q) \Phi^{'(F)}_0(p) \, .
\label{disc_A_forw}
\eeq

%In the saddle point computation for $n=0$ around $\nu=0$ the general
%nonforward expression is less reliable, depending on the energy, in the
%vicinity of the forward region. This is caused by the structure of the
%nonforward Green's function eq.(\ref{G2mom}), since the factor $N_h$ is
%quadratically suppressed in the neighborhood of $\nu=0$, and to the
%fact that in a small $l$ expansion one would need to shift the saddle
%point due to its dependence in $l$.
%We shall cure this problem in the approximation by interpolating the
%non forward results with the forward ones. 

Let us note that in the transition from the nonforward to forward
physics the perturbative QCD analysis is increasingly affected by the
long distance interactions
(contributions for all values of the impact parameter), since gluons are
massless.
This implies that some care is required to treat the general
non-forward expression in this limit. The saddle point
approximation for $n=0$ around $\nu=0$ leading to eq.(\ref{sad}) is
accurate only for $l$ down to values of the order of one. Strictly speaking,
performing the saddle point approximation of eq.(\ref{discAfinal}) 
in the region of very small $l$ one needs to shift the saddle point
which becomes $l$ dependent. 
In our numerical calculation we have used eq.(\ref{sad}) down to $|2
{\bf l}|^2=1/2\, M_Z^2$, and then we have extrapolated to the forward
value, eq.(\ref{disc_A_forw}).

%when computing its saddle point approximation ($n=0$ around $\nu=0$).
%Note also that the structure of the
%nonforward Green's function eq.(\ref{G2mom}): the factor $N_h$ is
%quadratically suppressed in the neighborhood of $\nu=0$, and in a small
%$l$ expansion one would need to shift the saddle
%point due to its dependence in $l$.
%We shall cure this problem in the approximation by interpolating the
%non-forward results with the forward ones.         

A more precise analysis, avoiding the use of the saddle point
approximation is of course possible but it is not needed for the order of
magnitude estimate which is pursued here. By the way few checks
have been done with an exact Green's function computation, confirming
such an expectation. 

Finally, in order to evaluate the elastic cross
sections we need to obtain the amplitude from the computed
discontinuity. Instead of using the full dispersion relation
technique, a simple evaluation compatible with the
approximation adopted is given by
\beq
A \sim \left( 1- i \frac{\pi}{2} \frac{\partial}{\partial y} \right)
\frac{1}{2}  Disc_s A \, ,
\label{disc2amp}
\eeq
as provided by the signature factor in the partial wave amplitude
analysis.
  
%%%%%%%%%%%%%%%%%%%%%%%%%%%%%%%%%%%%%%%%%%%%%%%%%%%%%%%%%%%%%%%%%%%%%%%
\section{The process $\gamma\gamma\to ZZ$ in the small angle limit}
We shall henceforth concentrate on a definite process, $\gamma\gamma\to
ZZ$, and compute for it a numerical estimate of the QCD contributions.
We remind that we are interested in the kinematical region of small
scattering angles, i.e. scattering close to the forward region,
so that one can do the calculation in the high-energy approximation
suited for the Regge limit:
terms which are suppressed by powers of $t/s$ can be
neglected, with a significant simplification in the computation.
We shall in fact use the results presented in section 2 and 3.
%%%%%%%%%%%%%%%%%%%%%%%%%%%%%%%%%%%%%%%%%%%%%%%%%%%%%%%%%%%%%%%%%%%%%
%%%%%%%%%%%%%%%%%%%%%%%%%%%%%%%%%%%%%%%%%%%%%%%%%%%%%%%%%%%%%%%%%%%%%%%%%%%%
\begin{figure}[t]
  \begin{center} %\hspace{0.3cm}
     \epsfig{file=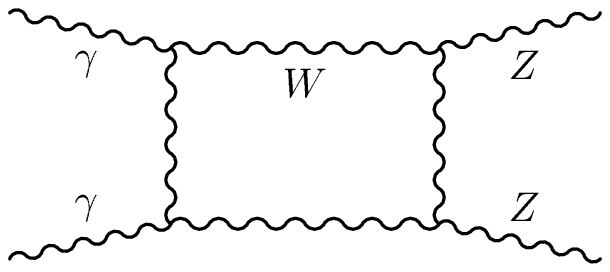,bbllx=210pt,bblly=640pt,bburx=398pt,bbury=722pt,clip=,width=6cm}
     \epsfig{file=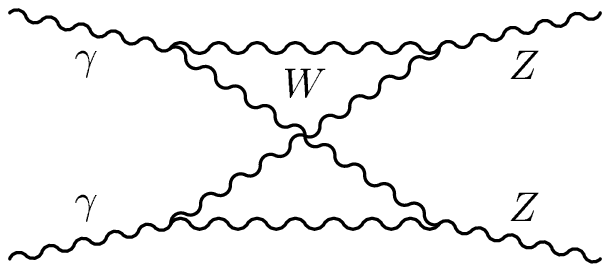,bbllx=210pt,bblly=640pt,bburx=398pt,bbury=722pt,clip=,width=6cm}
     \caption{The surviving contributions to the leading electroweak
     amplitude in the Regge limit for real $\gamma$ and transverse
     polarised Z.}
     \label{ggzzEW}
\end{center}
 \end{figure}
%%%%%%%%%%%%%%%%%%%%%%%%%%%%%%%%%%%%%%%%%%%%%%%%%%%%%%%%%%%%%%%%%%%%%%%%%%%%%

\subsection{The leading electroweak contribution}
Let us first analyze the pure electroweak contributions to the process.
This process starts at one loop and was calculated by Jikia
\cite{Jikia}.
In order to calculate the QCD corrections to $\gamma\gamma\to ZZ$ one
has to know the leading electroweak contribution at the amplitude
level, add to it the leading QCD amplitude and take the modulus squared.
The main contribution to the corrections will come from
the interference term.
In the high energy limit where $t$ is kept fixed most of the
contributions are suppressed by powers of $s$. The surviving
graphs are pictured in Fig. \ref{ggzzEW}. Also for these only the
helicity conserving amplitudes survive, with both of the bosons
transverse polarised. One may confirm this by taking the limit $s$ to
infinity and keeping $t$ fixed within the expressions of \cite{Jikia}.

In order to simplify the calculation we compute first the
$s$-channel discontinuity, i.e. the $s$-channel internal $W$ lines
are taken on-shell. Here the $s$-channel discontinuity coincides
with twice the imaginary part of the amplitude. With the same
assignment of the momenta already used in the last sections the
imaginary part of the box contribution reads:
\begin{equation}
  \label{eq:integ}
  {\rm Im}\, A^{(1)}_{\gamma\gamma\to ZZ}=\frac 12 \int \frac{d^4k}{(2\pi)^2}
 \, \delta [ (q-k)^2-M_W^2]\; \delta [ (p+k)^2-M_W^2]\,
 A^{(0)}_{\gamma\gamma\to WW} A^{(0)\dagger}_{WW\to ZZ} \, ,
\end{equation}
where a summation over the W boson helicities is implied. Performing
the loop integration and retaining only the leading power of $s$ we get: 
\begin{equation}
  \label{eq:imbox}
  {\rm Im}\, A_{Box}=s \, \alpha_{em}^2 \,\frac{c_w^2}{s_w^2}\,
  \frac{8\pi}{\sqrt{t^2-4tM_W^2}}\; \ln \left[ \left(
  \frac{-t+\sqrt{t^2-4tM_W^2}}{-t-\sqrt{t^2-4tM_W^2}} \right)^2
  \right]\, , 
\end{equation}
where $|2 {\bf l}|^2\approx -t$ was used.
Equation (\ref{eq:imbox}) yields immediately the real part by just noting
that $\ln (-s)=\ln s -i\pi$. Thus the reconstructed (leading) full box 
amplitude reads:
\begin{equation}
  \label{eq:box}
  A_{Box}=\frac{-s}{\pi} \ln \left( \frac{s}{t} \right) \,
  \alpha_{em}^2 \,\frac{c_w^2}{s_w^2}\,
  \frac{8\pi}{\sqrt{t^2-4tM_W^2}}\; \ln \left[ \left(
  \frac{-t+\sqrt{t^2-4tM_W^2}}{-t-\sqrt{t^2-4tM_W^2}} \right)^2
  \right].
\end{equation}
The amplitude of the crossed diagram is obtained from the box
contribution (\ref{eq:box}) by replacing $s$ by $u \approx -s$.
Adding the box and crossed contributions, at the end one
is left with the imaginary part of the box diagram:
\begin{equation}
  \label{eq:ew}
  A_{EW}^{++++}=A_{EW}^{+-+-}=i \; {\rm Im}\, A_{Box}
\end{equation}
Thus in the Regge limit the leading order electroweak contribution is 
purely imaginary  ($s\gg -t$), and the cross section is constant in
$s$. Considering the limiting behavior of eq.(\ref{eq:ew}) in the region 
$s\gg -t\gg M_W^2$ and for the case $t=0$ one obtains the expressions 
presented in \cite{Jikia}.

\subsection{QCD corrections}
The amplitude of this process in the high energy limit is illustrated
in Fig.~\ref{bfklampl}. The ingredients needed have been discussed in
sections 2 and 3. To proceed with an explicit calculation we have to specify
the couplings ${\it C_f}$, which depend on the external bosons, and
the virtualities appearing in the impact factors given in equations
(\ref{TT}) and (\ref{TL}).
The starting expressions are for massive quarks. For
practical purposes only the top quark mass has to be taken into
account because of the large Z mass.
The integrals which correspond to the quark loop
have to be performed. The last integration we compute is the scalar product
of the impact factors with the BFKL states
eq.(\ref{conf_repr_phi}) and (\ref{forw_conf_repr_phi}).

We calculate this process for all external bosons
on-shell. In such a case we have $Q_+^2$ identically zero and
$Q_-^2=-M_Z^2$.
The coupling constants ${\it C_f}$ - the product of the
$\gamma$-quark and Z-quark couplings - are given by:
\begin{equation}
  \label{eq:coupling}
  {\it C_f}=\frac{4 \pi \alpha_{em}}{2 \sin \theta_w \cos
    \theta_w}\left\{ q_f T_f^3  - 2 \, q_f^2 \sin^2\theta_w \right\}\, . 
\end{equation}

First we consider the scattering of the photons into transverse Z bosons.
To proceed we have to solve the remaining integrations for the
impact factor $\Phi_{TT}$ in eq.(\ref{TT}). In order to compute the
integral over the two dimensional euclidean vector ${\bf u}$ we
employ a standard Feynman parametrization and rewrite eq.(\ref{TT}) as:
\begin{eqnarray}
  \label{eq:feyn}
&& \hspace{-0.7cm} \Phi^{(ij)}_{TT}(q,k,l) = \alpha_s\,
  \sqrt{N_c^2-1}\;
\sum_f {\it C}_f\,\sqrt{4\pi}\int_0^1 d\alpha \int
\frac{d^2{\bf u}}{(2\pi)^2} \int_0^1 dx\times \\
&& \hspace{-0cm}\nonumber\times \left\{ \frac{4\al
    (1-\al)[\epsbo_i\cdot{\bf u}\;\epsbo_j^*\cdot{\bf
      u}-x(1-x)\epsbo_i\cdot{\bf p_a}\;\epsbo_j^*\cdot{\bf p_a}
    ]-(u^2-x(1-x)p_a^2+m_f^2)(\epsbo_i\cdot
  \epsbo_j^*)}{[u^2+x(1-x)p_a^2-x m+m^2_f]^2}-\right. \\
&&\nonumber - \left. \frac{4\al (1-\al)[\epsbo_i\cdot{\bf
      u}\;\epsbo_j^*\cdot{\bf u}-x(1-x)\epsbo_i\cdot{\bf
      p_b}\;\epsbo_j^*\cdot{\bf p_b} ]-(u^2-x(1-x)p_b^2+m_f^2)(\epsbo_i\cdot
  \epsbo_j^*)}{[u^2+x(1-x)p_b^2-x m+m^2_f]^2}
\right\} \, .
\end{eqnarray}
Here we have defined ${\bf p_a}=-{\bf k}-(1-2\alpha){\bf l}\; , \; {\bf
  p_b}=2\alpha {\bf l}$ and $m=\al (1-\al)(M_Z^2+i\epsilon)$. 
Strictly speaking, at this stage of the calculation $m$ is taken
negative and subsequently, after having performed the integrations,
analytically continued to positive values obtaining the above definition.    
The first fraction corresponds to the two diagrams where one of the
  gluons is attached to the quark and the other to the antiquark. The
  second fraction corresponds to the two diagrams where the two
  gluons are attached to the same (anti)quark.
Each of them would lead to a divergent integral but, or course, the sum in
 eq.(\ref{eq:feyn}) gives a finite result.

On performing the ${\bf u}$ integration we end up with:
\begin{eqnarray}
  \label{eq:xalphaTT}
\hspace{-0cm} \nonumber\Phi^{(ij)}_{TT}(q,k,l) &&=\alpha_s\, \sqrt{N_c^2-1}\; 
\sum_f {\it C}_f\,\sqrt{4\pi}\int_0^1 \frac{d\alpha}{2\pi} \int_0^1 dx\;
\bigg\{ -4\alpha(1-\alpha)\, x(1-x)  \times \\ 
&&\nonumber \hspace{0cm} \times\left( \frac{\epsbo_i\cdot{\bf p_a}\;
    \epsbo_j^*\cdot{\bf p_a}}{x(1-x)p_a^2-xm+m_f^2}-
\frac{\epsbo_i\cdot{\bf p_b}\;
    \epsbo_j^*\cdot{\bf p_b}}{x(1-x)p_b^2-xm+m_f^2} \right)+ \\
&& \hspace{0cm} \nonumber +  \left. \epsbo_i \cdot\epsbo_j^*\,
\left( [\alpha^2+(1-\alpha)^2] \ln\left[
    \frac{x(1-x)p_a^2-xm+m_f^2}{x(1-x)p_b^2-xm+m_f^2}\right] +\right.\right.\\ 
&& \hspace{0cm}+ \left.\left. 
    \frac{ x(1-x) p_a^2-m_f^2}{x(1-x)p_a^2-xm+m_f^2}-
\frac{ x(1-x) p_b^2-m_f^2}{x(1-x)p_b^2-xm+m_f^2}\right) \right\}.
\end{eqnarray}

The computation of the $x$-integration is straightforward. 
Due to the lenght of the result we present here only the limit $m_f
\to 0$ which preserves a significant simplification compared to the
massive case. The corresponding helicity conserving result 
with massive fermions is presented in the appendix.
After the integration in the Feynman parameter variable,
one obtains in the zero quark mass limit for the separate helicity states:
\begin{eqnarray}
  \label{eq:alphaTT}
\hspace{-0cm} \Phi^{(++)}_{TT}(q,k,l) && =\alpha_s\, \sqrt{N_c^2-1}\; 
\sum_f {\it C}_f\, 
\sqrt{4\pi}\int_0^1 \frac{d\alpha}{2\pi}\; [\alpha^2+(1-\alpha)^2]\;
\ln\left[ \frac{p_a^2-m}{p_b^2-m}\right]\\
\hspace{-0cm} \Phi^{(--)}_{TT}(q,k,l) && = \Phi^{(++)}_{TT}(q,k,l) \\ 
\hspace{-0cm} \Phi^{(+-)}_{TT}(q,k,l) && = i \,\alpha_s\,
\sqrt{N_c^2-1}\sum_f {\it C}_f\, 
\sqrt{4\pi}\int_0^1 \frac{d\alpha}{2\pi} \;4\alpha(1-\alpha)\times\\
&& \hspace{0cm}\times\nonumber\left\{ \frac{p_a^x p_a^y}{p_a^2}
\left( 1+\frac m{p_a^2}\ln\left[ \frac{p_a^2-m}{-m}\right]\right)
 -\frac{p_b^x p_b^y}{p_b^2} \left(
1+\frac m{p_b^2}\ln\left[ \frac{p_b^2-m}{-m}\right]\right) \right\}\\
\hspace{-0cm} \Phi^{(-+)}_{TT}(q,k,l) && =- \Phi^{(+-)}_{TT}(q,k,l) \, ,
\end{eqnarray}
with $(p^x,p^y)={\bf p}$. For the helicity conserving parts in the
massless quark limit the remaining $\al$ integration was
also performed analytically and is presented in the appendix.
For the result with massive quarks a numerical integration has been
performed.

In the same way one can calculate the scattering of real photons into
longitudinal Z-bosons and obtain (in the $m_f\to 0$ limit):
\begin{eqnarray}
  \label{eq:alphaTL}
\nonumber  \Phi^{(i)}_{TL}(q,k,l) && = \alpha_s\,
\sqrt{N_c^2-1}\sum_f{\it C}_f\, 2 
\sqrt{4\pi}\; M_Z \int_0^1 \frac{d\alpha}{2\pi}\; \;
\al (1-\al)(1-2\al)\times \\
&& \times \left\{ \frac{\epsbo_i\cdot{\bf p_b}}{p_b^2}
\ln\left[ \frac{p_b^2-m}{-m}\right]- \frac{\epsbo_i\cdot{\bf p_a}}{p_a^2}
\ln\left[ \frac{p_a^2-m}{-m}\right]\right\}.
\end{eqnarray}

In order to compute the two-gluon or the BFKL amplitude one has to
substitute these expressions into eq.(\ref{2glampl}) or eqs. 
(\ref{conf_repr_phi}, \ref{forw_conf_repr_phi}).

%%%%%%%%%%%%%%%%%%%%%%%%%%%%%%%%%%%%%%%%%%%%%%%%%%%%%%%%%%%%
\section{Numerical results}
In this section we present numerical results on the QCD
corrections for ZZ pair production in $\gamma\gamma$ collisions. 
Here we consider full circular polarization of the incoming photons.
The relative QCD contributions are presented for
transverse Z bosons only, because in the high-energy approximation
the helicity changing parts of the electroweak amplitude are
supressed by powers of $s$ and thus are negligible. 
Nevertheless we shall display the helicity changing part of the BFKL
amplitude as the pure QCD contribution alone. The parameters
$\al_W=\al/s_W^2$, $\al =1/128$, $m_Z=91.2$ GeV, $m_t=174.3$ GeV 
and $\alpha_s (s_0)$
have been used throughout the numerical computations. 
We have been neglecting all quark masses apart
from the top quark mass in the numerical calculation, since they are
neglible compared to the Z boson mass. The inclusion of the top quark mass
reduces the QCD amplitude by 25\%. 

In terms of the amplitudes
(\ref{eq:ew},\ref{discAfinal},\ref{disc2amp})
the cross section reads:
\begin{equation}
  \label{eq:cross}
  \frac{d\sigma}{dp_T^2}=\frac1{16\pi s^2}|A_{EW}+A_{QCD}|^2\,
\end{equation}
where $p_T=|2\,{\bf l}|$ is the exchanged transverse momenta.
The angular distribution of the cross section is obtained by the
relation:
\begin{equation}
  \label{eq:ang}
  dt=\frac s2 \beta\;\, d\cos\theta \, ,
\end{equation}
with $\beta =\sqrt{1-4M_Z^2/s}$.
For small scattering angles in the high-energy approximation the
two incoming and two outgoing particles do not 'communicate' with each other 
because of the huge separation in rapidity. 
As a result, the amplitude with the helicity 
$++\to ++$ is equal to the amplitude with helicity $ +-\to +-$ etc. 
Since the helicity conserving amplitudes completely dominate the cross section,
after the summation of outgoing helicities the
result is the same for all the possibilities of incoming helicities,
i.e. $ ++\to \rm{TT}$ is equal to $+-\to \rm{TT}$ etc. In the
following $\gamma\gamma\to Z_T Z_T$ stands for all these possibilities
as well as for unpolarised scattering.

In Fig. 4 the angular distribution of the leading electroweak
contribution for transverse Z pair
production for polarised $\gamma\gamma$ collision at
$\sqrt{s_{\gamma \gamma}}= 1$ TeV is given. 
The solid line is the result of the calculation in the high-energy
approximation, presented in section 4.1. 
The dotted line is the result of the exact calculation, presented also
in \cite{Jikia} for $\sqrt{s_{\gamma \gamma}}= 500$ GeV. 
For $\cos\theta \approx 0.9$ the result of the high-energy approximation has
a deviation of $\sim$ 15\% from the
exact calculation, for smaller scattering angles the deviation is decreasing.  
With rising center of mass energy the accuracy of the Regge
calculation is continously increasing.

Let us consider the QCD corrections.
In the Born analysis, from the naive counting of the couplings (see
Fig.~\ref{2gl}),
one expects in the interference term a suppression,
with respect to the one loop EW case, due to the factor
$\alpha_s^2$ and an additional suppression of a factor $\approx 4$
due to the difference in the EW couplings. This would lead to
corrections of less than a percent.
At high energies in the Regge limit we expect an enhancement due to
the appearance of some large logarithms, which may be resummed in the
BFKL scheme. We stress that the resummed leading log QCD
corrections at the lower energies we consider are overshooting what is
expected from the real contribution, nevertheless they provide
a first estimate.
We remind, as discussed
previously, that the BFKL resummation is evaluated in the saddle point
approximation.
Fig. 5 displays the differential cross section with respect to the
exchanged transverse momenta at a center of mass energy of
$\sqrt{s_{\gamma \gamma}}= 1$ TeV.      
The solid line is the pure electroweak cross section, and the dashed
line is the one with the QCD corrections included,
eq.(\ref{eq:cross}). The cross section
has a slight rise approaching the forward region.
This slight rise comes from the electroweak part since it completely
dominates the amplitude.
As in the following it will be visible, for the QCD part of the cross
section the rise approaching the forward
region is much stronger. Therefore the QCD corrections are enhanced
close to the forward region. 
%%%%%%%%%%%%%%%%%%%%%%%%%%%%%%%%%%%%%%%%%%%%%%%%%%%%%%%%%%%%%%%%%%%%%%%%%%
\begin{figure} 
\hspace{1cm}\label{fig:EW}
\epsfig{width=13cm,file=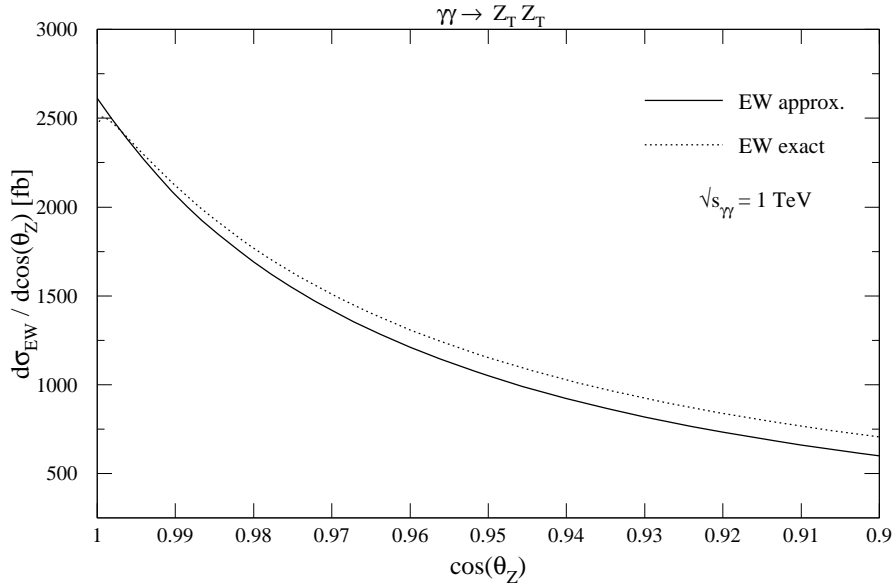}
\caption{Angular distribution of the leading electroweak contribution
  for transverse Z pair production in
  polarised $\gamma\gamma$ collision  
  at $\sqrt{s}=1\; {\rm TeV}$. The solid line is the result of the 
  high-energy approximation, the dotted line is the result of
  the exact calculation.}
\end{figure} 
%%%%%%%%%%%%%%%%%%%%%%%%%%%%%%%%%%%%%%%%%%%%%%%%%%%%%%%%%%%%%%%%%%%%%%%%%%%%
%%%%%%%%%%%%%%%%%%%%%%%%%%%%%%%%%%%%%%%%%%%%%%%%%%%%%%%%%%%%%%%%%%%%%%%%%%
\begin{figure} 
\hspace{1cm}\label{fig:cross} 
\epsfig{width=13cm,file=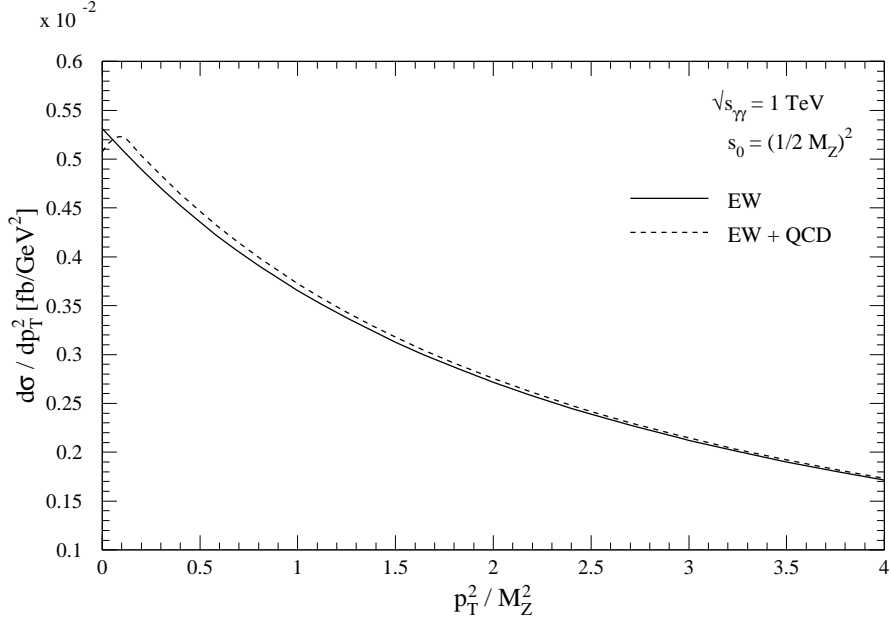}  
\caption{The pure EW differential cross section (solid line) versus
  the exchanged transverse momenta at
  $\sqrt{s}=1\; {\rm TeV}$ and the one which include the QCD
  corrections (dashed line). For the BFKL contribution the scale
  $s_0=(1/2 M_Z)^2$ was used.}
\end{figure} 
%%%%%%%%%%%%%%%%%%%%%%%%%%%%%%%%%%%%%%%%%%%%%%%%%%%%%%%%%%%%%%%%%%%%%%%%%%
%%%%%%%%%%%%%%%%%%%%%%%%%%%%%%%%%%%%%%%%%%%%%%%%%%%%%%%%%%%%%
\begin{figure}\label{fig:corr} 
\begin{center} 
\hspace{-1cm}
\epsfig{width=16cm,file=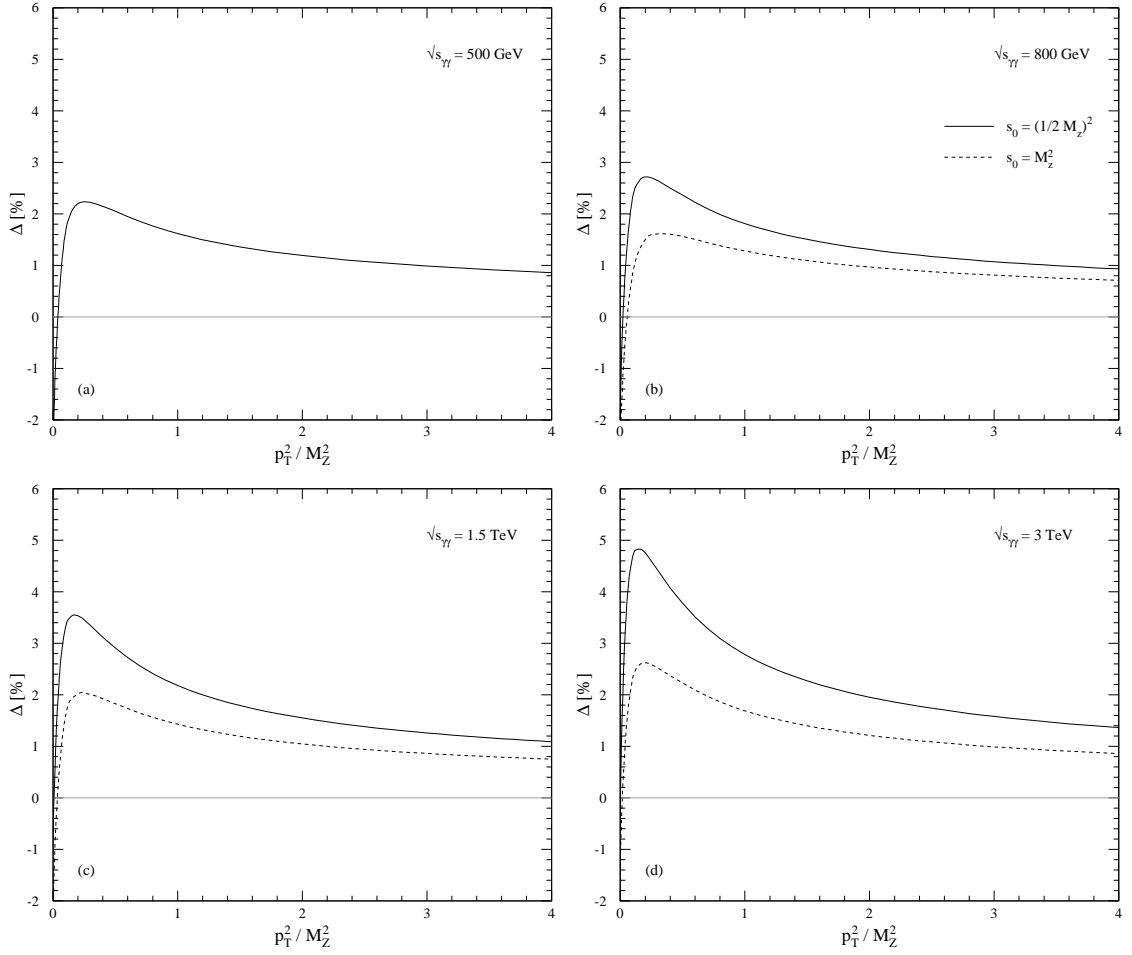} %\vspace{-1cm}
\end{center}
\caption{QCD corrections to the differential cross section relative to
  the pure EW contribution for different scales $s_0$ and center of
  mass energies $\sqrt s$. 
  The relative correction is defined as
  $\Delta= (\frac{d\sigma_{QCD+EW}}{dp_T^2} -
  \frac{d\sigma_{EW}}{dp_T^2})/\frac{d\sigma_{EW}}{dp_T^2}$.}
\end{figure}        
%%%%%%%%%%%%%%%%%%%%%%%%%%%%%%%%%%%%%%%%%%%%%%%%%%%%%%%%%%%%%%%%%%%%%%%%
%%%%%%%%%%%%%%%%%%%%%%%%%%%%%%%%%%%%%%%%%%%%%%%%%%%%%%%%%%%%%
\begin{figure}\label{fig:intcorr} 
\hspace{.5cm}  
\epsfig{width=14cm,file=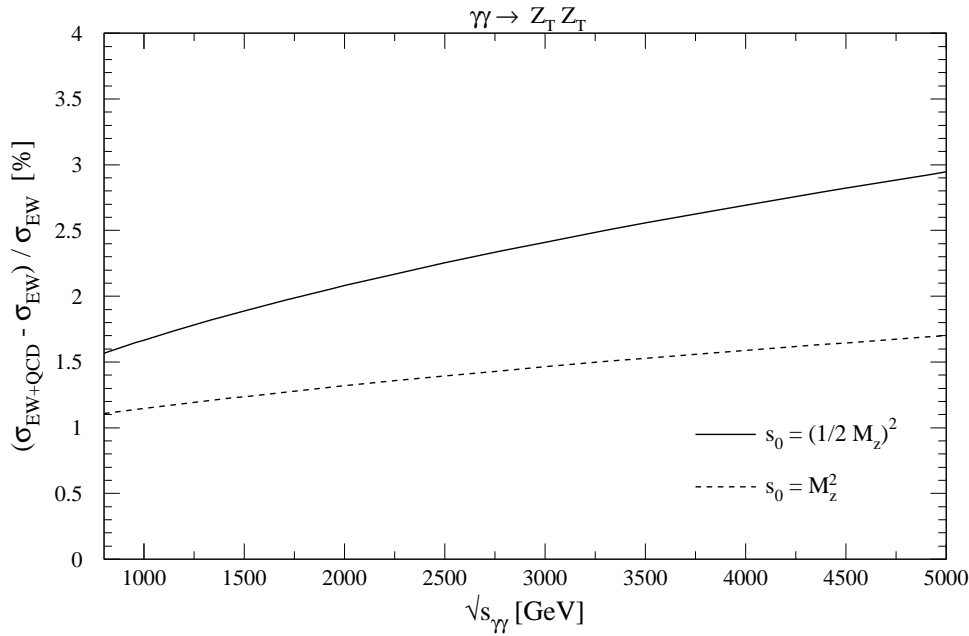}   
\caption{The QCD corrections to the integrated cross section relative
  to the pure EW contribution for different scales. The integration
  region was choosen as $1/4\, M_Z^2 < p_T < 4\, M_Z^2$.}
\end{figure}        
%%%%%%%%%%%%%%%%%%%%%%%%%%%%%%%%%%%%%%%%%%%%%%%%%%%%%%%%%%%%%%%%%%%%%%
%%%%%%%%%%%%%%%%%%%%%%%%%%%%%%%%%%%%%%%%%%%%%%%%%%%%%%%%%%%%%
\begin{figure}[t]
\label{fig:diag}
\hspace{.5cm}
\epsfig{width=14cm,file=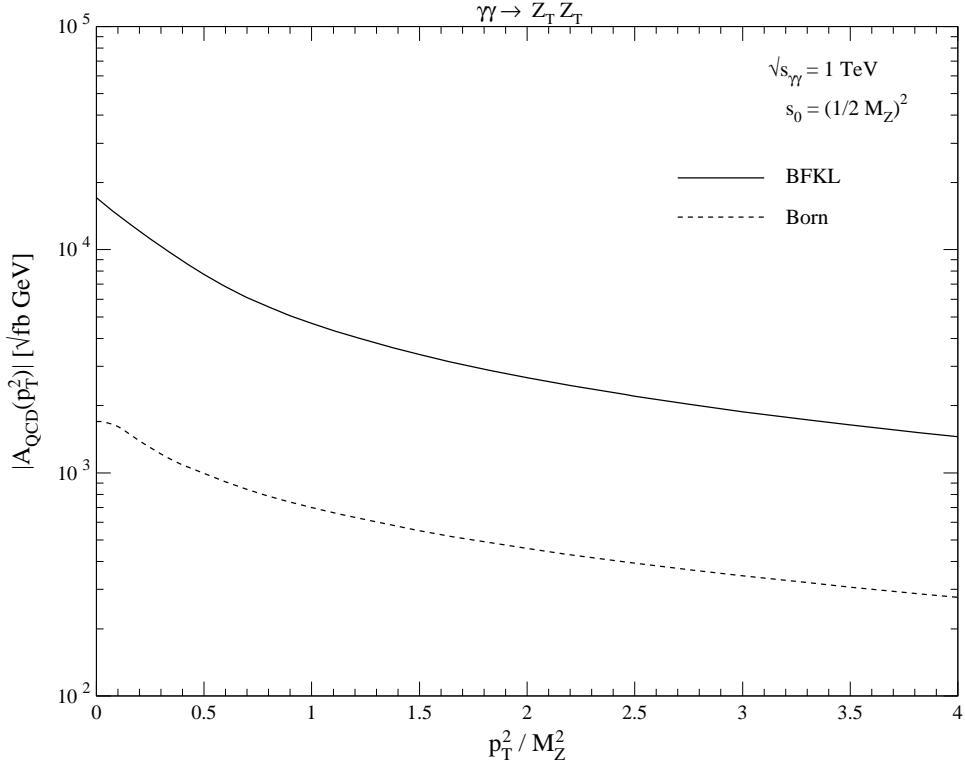}
\caption{The pure QCD part of the amplitude versus the
  exchanged transverse momenta. The solid line is the BFKL
  contribution and the
  dashed line is the QCD Born approximation (two-gluon amplitude).}
\end{figure}        
%%%%%%%%%%%%%%%%%%%%%%%%%%%%%%%%%%%%%%%%%%%%%%%%%%%%%%%%%%%%
\begin{figure}\label{fig:nondiag}
\hspace{.5cm} 
\epsfig{width=14cm,file=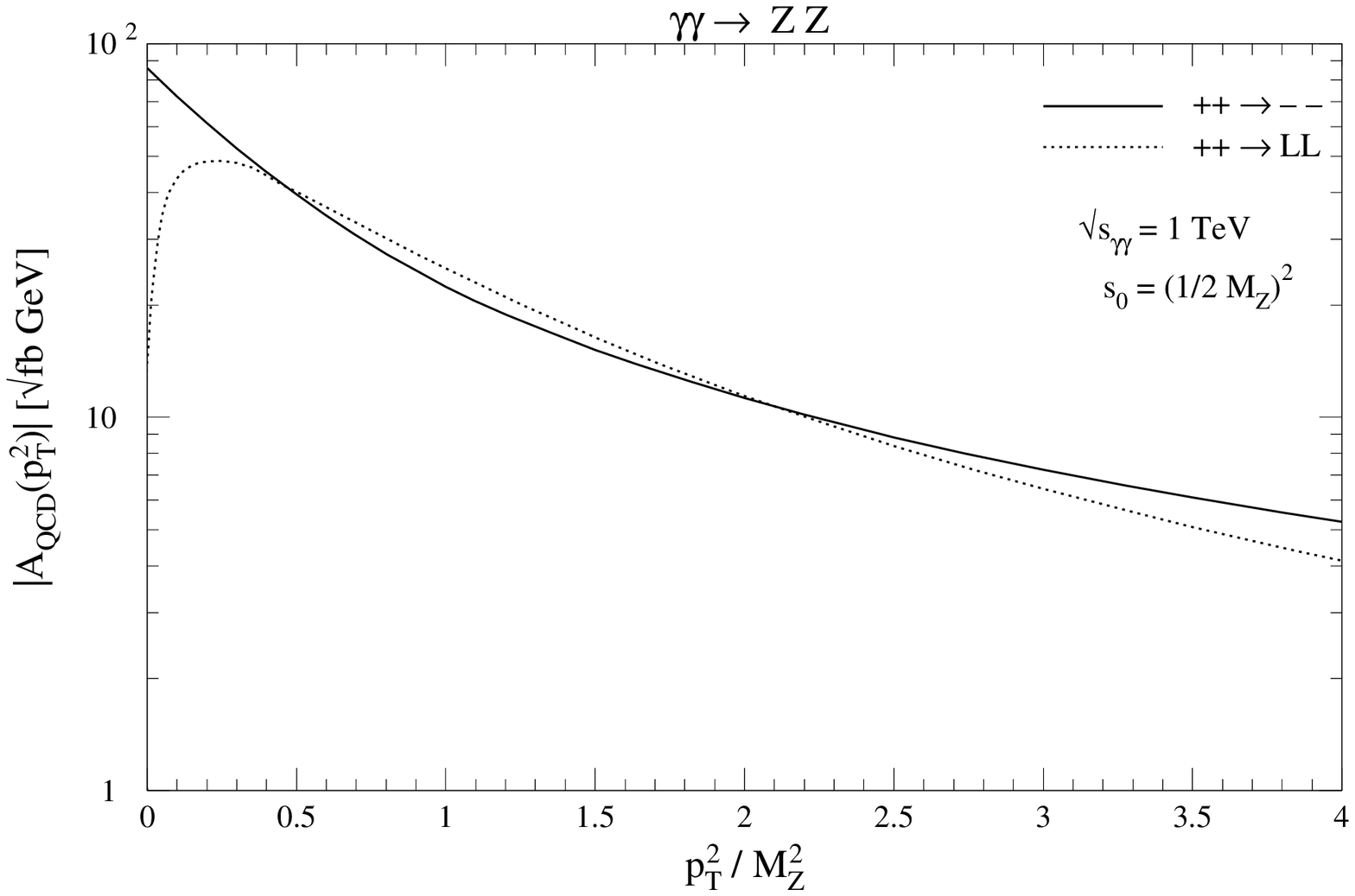}
\caption{The same as Fig.~\ref{fig:diag} for the helicity changing parts of
  the BFKL contribution.}
\end{figure}        
%%%%%%%%%%%%%%%%%%%%%%%%%%%%%%%%%%%%%%%%%%%%%%%%%%%%%%%%%%%

This behavior is shown in Fig.~$6$, where
the QCD corrections to the differential
cross section relative to the pure electroweak contribution 
at center of mass energies of 500 GeV (a), 800 GeV (b), 1.5 TeV (c) and 3
TeV (d) are shown as functions of $p_T^2/M_Z^2$.
At these energies $p_T^2/M_Z^2=4$ corresponds to values of $\cos\theta$
(where $\theta$ is the scattering angle) of $0.75$ (a), $0.90$ (b),
$0.97$ (c) and $0.99$ (d) respectively.
This relative correction is defined as
\begin{equation}\label{eq:corr}
\Delta=\left(\frac{d\sigma_{QCD+EW}}{dp_T^2} - 
  \frac{d\sigma_{EW}}{dp_T^2}\right)\left/
  \frac{d\sigma_{EW}}{dp_T^2}\right. \, .
\end{equation}
Since for leading order BFKL the scale is not fixed, there is an 
uncertainty which affects the magnitude of the QCD corrections. The solid
lines in Fig.~$6$ are results for the scale $s_0=(1/2\;
M_Z)^2$ while the dashed lines correspond to $s_0=M_Z^2$.
At the scale $s_0=M_Z^2$ the minimal center of mass energy $\sqrt s$,
for which the saddle point approximation for the leading logarithmic calculations gives a reliable result, is 800 GeV. 
Corrections resulting from leading order BFKL contribution are at the 
percent level. For higher $p_T$ they are about one
percent and  approaching the forward region (smaller $p_T$ values)
they rise and lead to corrections of a few percent. 
Close to the forward region the sign of the corrections is changing as
a result of the fact that for small ${\bf l}$ the imaginary part
of $\Phi_h$ (eq.20) gets an enhancement. 
Fig.~$6$ (a-d) summarizes the dependencies of the corrections
resulting from leading order BFKL calculations with respect to
the center of mass energy $\sqrt s$, the exchanged transverse
momenta $p_T$ and the scale $s_0$.
For rising center of mass energy the corrections increase. 
The solid lines ($s_0=(1/2\; M_Z)^2$) in Fig.~$6$(a) and 
(d) show that the corrections may rise in this regime by a factor of 2 if the
energy increases from 500 GeV to 3 TeV. The scale $s_0$ implicates
also a strong dependence. In fact close to the forward region the corrections
also grow by a factor of 2 if the scale is decreasing from $s_0=M_Z^2$ to 
$s_0=(1/2\; M_Z)^2$. The biggest dependence is coming from the
exchanged momenta or equivalently from the scattering angle. The
corrections have a strong rise with decreasing scattering angle or
$p_T$ and change sign close to the forward region. 

In Fig. 7 relative corrections to the integrated cross
section $(\sigma_{EW+QCD}-\sigma_{EW})/\sigma_{EW}$ are plotted as a 
function of the center of mass energy. 
These are also at the order of percent level.
The corrections rise with increasing center of mass energy. The
magnitude of the corrections is again scale dependent and the rise is
stronger for smaller scales, as expected.
Since the calculations are done in the Regge limit we integrated
eq.(\ref{eq:cross}) in the region of $1/4\, M_Z^2 < p_T < 4\, M_Z^2$.
An integration down to $p_T=0$ would change the result by less then
0.1 \%, since the sign of the corrections changes very close to the
forward region, (see Fig. 6). 
%Anyway this region is excluded by the experimental reality.
If one integrates over all scattering angles the corrections
become less important, but since the main contribution of the cross
section is coming from small scattering angles the result would not be
affected strongly. 

In the following figures some properties of the pure QCD part of the
amplitude are discussed. 
In Fig.~8 the absolute value of the QCD amplitudes is
displayed for the helicity conserving case.
%Since the helicity conserving parts completely
%dominate the cross section we would get the same result for any
%photon polarisation with the Z boson polarisation summed.
The solid line is the BFKL (resummed) contribution while 
the dashed line is the Born (two-gluon) contribution.
At $\sqrt s=1$ TeV the gluon resummation is enhancing the pure
QCD Born amplitude by one order of magnitude.
The same enhancement is visible in the QCD corrections to the 
electroweak cross section. 
Both amplitudes are rising with decreasing $p_T$, whereas for higher
energies the rise of the resummed amplitude will be stronger as the Born one.
%Furthermore the rise of the amplitude approaching the
%forward limit is slightly stronger in the resummed case. It is rising by
%one order of magnitude from $p_T^2/M_Z^2=4$ to the forward region.  
%The dashed line in Fig.~8 is the BFKL contribution and the solid line
%is the same up to $p_T^2/M_Z^2=0.5$ and then extrapolated to the
%forward value calculated with eq.(\ref{disc_A_forw}), according to
%what already discussed in the second last paragraph of section $3$.
%The dashed line in Fig.~8 is the non-forward BFKL contribution, which is
%overshooting according to what already discussed in the second last
%paragraph of section 3. The solid line
%is the same up to $p_T^2/M_Z^2=0.5$ and then extrapolated to the
%forward value, calculated with eq.(\ref{disc_A_forw}), giving the reliable
%result for the amplitude.     

In Fig. 9 the absolute value of the two helicity changing parts of the
amplitude at $\sqrt{s}=1$TeV are plotted. The most noticable property
is that the helicity changing parts are suppressed by more than two
orders of magnitude with respect to the helicity conserving part. The $++\to --$
cross section has, apart from its magnitude, properties similar to $++\to
++$, for example its slope as a function of $p_T$.
The $++\to LL$ part is also rising when $p_T$ is getting smaller but
in the region of $p_T=1/2\, M_Z$ it has
an extremum and is steeply going to zero for $p_T=0$. This behavior
comes from eq.(\ref{TL}), where the integrand is antisymmetric
in $\alpha \leftrightarrow (1-\alpha)$ for ${\bf l}=0$.       
%%%%%%%%%%%%%%%%%%%%%%%%%%%%%%%%%%%%%%%%%%%%%%%%%%%%%%%%%%%%%%%%%%%%
\section{Conclusions}

In this paper we have presented formulae for the leading QCD corrections to vector boson
scattering at small scattering angles. These emerge since, like the
photon, all vector
bosons fluctuate into quark-antiquark pairs, which are strongly
interacting colour dipoles. The nonforward boson impact factors describe the
probability that the boson fluctuates into a quark-antiquark
dipole. First the Born and after the leading logarithmic resummed
(BFKL) contributions have been investigated for an arbitrary process of
elastic scattering of vector bosons. 

In particular the process $\gamma\gamma\to ZZ$ has been considered in detail. 
Both, the electroweak and the QCD parts
of the amplitude have been calculated in the high-energy approximation. 
The results are displayed for the differential and integrated
cross section, comparing QCD corrections relative to the one loop
elektroweak calculation.
We discussed some properties of the pure QCD part of
the cross section as well, including also nonleading contributions.
The general result of the numerical calculation is that at $s=O(1$~TeV) the QCD
corrections affect the cross section of the process by a few percents.
The corrections are smaller for larger $p_T$ values and rise with
falling $p_T$.
Moreover they rise with the center of mass energy as expected.
Since the leading log BFKL contribution ends up with a noticable scale
dependence this results should not be regarded as a precision calculation.
Therefore, we interpret this result as an indication that, in
particular in the small angle region,
possible deviations from EW calculations cannot be interpreted as
signals of new physics without taking into account the QCD corrections.
%warning against the temptation
%to interpretate percent level deviations from electroweak
The
same reasoning is valid for the integrated cross sections, since the
biggest contribution to the cross section is coming from scattering at
small angles.

\bigskip
\centerline{\bf Acknowledgements}

We wish to thank J. Bartels for many helpful discussions.
Furthermore we wish to thank G. Chachamis for giving us data
regarding exact EW calculations, and S. Dittmaier for helpful discussions. 
K. P. is supported by the \textit{Graduiertenkolleg}
"Zuk\"unftige Entwicklungen in der Teilchenphysik".
G.P. Vacca thanks J. Bartels and the II. Institut f\"ur Theoretische Physik
for the warm hospitality.

%%%%%%%%%%%%%%%%%%%%%%%%%%%%%%%%%%%%%%%%%%%%%%%%%%%%%%%%%%%%%%%%%%%%%%%%%%%%
\section{Appendix}

In this appendix we present further formulas for the impact factor
calculated in chapter 4.2. 
 
In the case of massive fermions eq.(\ref{eq:alphaTT}) has the form:
\begin{eqnarray}\nonumber
\Phi^{(++)}_{TT}(q,k,l) &=& 
\alpha_s\, \sqrt{N_c^2-1}\; 
\sum_f {\it C}_f\, 
 \frac 1{\sqrt\pi}
\int_0^1 d\alpha \, \bigg\{ [ \alpha^2+\left( 1 -
        \alpha  \right)^2 ]  \times \\ \nonumber &&\hspace{-.3cm}\times
  \left[ \frac{p_a^2-m+2\,m_f^2}{2\,\Delta_a} \,
       \left( \log \left[ \frac{-m -p_a^2 -\Delta_a}{-m -p_a^2
             +\Delta_a}\right]  -  
         \log \left[ \frac{-m+p_a^2 - \Delta_a}{-m+p_a^2 + \Delta_a}\right] 
  \right. \right)  
\\ \nonumber &&\hspace{-.3cm}-\left. \frac{p_b^2-m+2\,m_f^2 }{2\,\Delta_b}  
       \left( \log \left[ \frac{-m - p_b^2 - \Delta_b}{-m - p_b^2 + \Delta_b}\right] -
         \log \left[ \frac{-m + p_b^2 - \Delta_b}{-m + p_b^2 +
             \Delta_b}\right]\right) \right] \\ \nonumber &&\hspace{-.3cm} -   
\frac{m_f^2}{\Delta_a}  \left( \log \left[ \frac{-m -p_a^2 -\Delta_a}{-m -p_a^2
             +\Delta_a}\right]  -
         \log \left[ \frac{-m + p_a^2 - \Delta_a}{-m + p_a^2 + \Delta_a}\right] 
  \right) \\
        &&\hspace{-.3cm}+ \left.
    \frac{m_f^2}{\Delta_b}\left( \log \left[ \frac{-m - p_b^2 -
        \Delta_b}{-m - p_b^2 + \Delta_b}\right] -
         \log \left[ \frac{-m + p_b^2 - \Delta_b}{-m + p_b^2 +
\Delta_b}\right]\right)\right\}  \, ,
\end{eqnarray}
while eq.(\ref{eq:alphaTL}) becomes:

\begin{eqnarray} 
\Phi^{(i)}_{TL}(q,k,l) &=& 
\alpha_s\, \sqrt{N_c^2-1}\; 
\sum_f {\it C}_f\, 
 \frac 2{\sqrt\pi} M_Z
\int_0^1 d\alpha \,   \bigg\{ \al(1-\al)(1-2\al) \times \\ \nonumber
&& \hspace{-2.15cm} \left[ \frac{\epsbo_i\cdot{\bf p_b}}{2\, p_b^2} \left(
    \frac{p_b^2-m}{\Delta_b} \left( \log \left[ \frac{-m - p_b^2 -
          \Delta_b}{-m - p_b^2 + \Delta_b}\right] - \log \left[
        \frac{-m + p_b^2 - \Delta_b}{-m + p_b^2 + \Delta_b}\right]-
\log \left[ \frac{m_f^2-m}{m_f^2}\right]
\right)\right) \right]
\\ \nonumber
&& \hspace{-2.5cm} - \left[ \frac{\epsbo_i\cdot{\bf p_a}}{2\, p_a^2} \left(
    \frac{p_a^2-m}{\Delta_a} \left( \log \left[ \frac{-m - p_a^2 -
          \Delta_a}{-m - p_a^2 + \Delta_a}\right] - \log \left[
        \frac{-m + p_a^2 - \Delta_a}{-m + p_a^2 + \Delta_a}\right]-
\log \left[ \frac{m_f^2-m}{m_f^2}\right]
\right)\right) \right]
\end{eqnarray}

with
\begin{eqnarray}\nonumber
  \Delta_a &=& \sqrt{(p_a^2-m)^2+4m_f^2\, p_a^2}\, ,\\
 \Delta_b &=& \sqrt{(p_b^2-m)^2+4m_f^2\, p_b^2}\, ,
\end{eqnarray}
and $p_a^2$, $p_b^2$ and $m$ was already defined in section 4.2. 
Taking the limit $m_f\to 0$ one obtains eq.(\ref{eq:alphaTT}) and
eq.(\ref{eq:alphaTL}) respectively.
This formulas were used for the top quark contribution,
where the remaining $\alpha$ integration was done numerically.

In the case of massless fermions for the helicity conserving part 
the $\alpha$ integration may be
performed analytically, giving the result: 
\begin{eqnarray}\nonumber
\Phi^{(++)}_{TT}(q,k,l) &=& 
\alpha_s\, \sqrt{N_c^2-1}\; 
\sum_f {\it C}_f\, 
 \frac 1{\sqrt\pi} \left\{ 2\left( x_1 - x_1^2 + x_2 - x_2^2 \right)
   + \right.
\\ \nonumber && \hspace{-.3cm} + 
  \left( 2 - 3\,x_1 + 3\,x_1^2 - 2\,x_1^3 \right) \log (1 - x_1) + 
 \left( 3\,x_1 - 3\,x_1^2 + 2\,x_1^3 \right)  
\log (-x_1) + \\ \nonumber && \hspace{-.3cm}+ 
\left( 2 - 3\,x_2 + 3\,x_2^2 - 2\,x_2^3 \right) \log
(1 - x_2) + 
  \left( 3\,x_2 - 3\,x_2^2 + 2\,x_2^3 \right) \log (-x_2) + 
 \\ \nonumber && \hspace{-.3cm}
+ \frac1{3 a^3} \left[ d\,\left( -3\,a^2 + 3\,a\,d+ 2\,d^2 \right)
  \,\log (\frac{d}{a}) + \right. 
 \\  &&  \hspace{-.3cm}+\left. \left.  
  \left( a + d \right) \left( 2\,a\,d - \left( 2\,a^2 + a\,d +
      2\,d^2 \right) \,\log (1+\frac da) \right)\right] \right\}\, , 
\end{eqnarray}

with
\begin{eqnarray} \nonumber
  a&=&4\, {\bf l}^2+M_Z^2+ i \epsilon \\ \nonumber
  b&=&-4({\bf l^2}+{\bf kl})-M_Z^2-i \epsilon \\ \nonumber
  c&=&({\bf k+l})^2 \\ \nonumber
  d&=&-M_Z^2- i \epsilon \\ 
  x_{1,2}&=&\frac{-b\pm \sqrt{b^2-4ac}}{2a} .
\end{eqnarray}

%%%%%%%%%%%%%%%%%%%%%%%%%%%%%%%%%%%%%%%%%%%%%%%%%%%%%%%%%%%%%%%%%%%%%%%%%%%%

\end{document}